\documentclass{IEEEtran}
\usepackage{cite}
\usepackage{amsmath,amssymb,amsfonts}
\usepackage{graphicx,subfigure,color}
\usepackage{textcomp,nicefrac}
\def\BibTeX{{\rm B\kern-.05em{\sc i\kern-.025em b}\kern-.08em
T\kern-.1667em\lower.7ex\hbox{E}\kern-.125emX}}
\usepackage{doi}
\usepackage[switch]{lineno}

\begin{document}
\title{Measurements of neutron fields in a wide energy range using multi-foil activation analysis}
\author{D. Chiesa, C. Cazzaniga,  M. Nastasi, M. Rebai, C. D. Frost, G. Gorini, S. Lilley, S. Pozzi, E. Previtali
\thanks{D. Chiesa, M. Nastasi, G. Gorini, S. Pozzi, and E. Previtali, are with the Dipartimento di Fisica, Università degli Studi di Milano - Bicocca and with the INFN - Sezione di Milano Bicocca, Milano I-20126, Italy (e-mail: davide.chiesa@mib.infn.it).}
\thanks{C. Cazzaniga, C. D. Frost, and S. Lilley are with the ISIS Facility, UKRI-STFC, Rutherford Appleton Laboratory, Didcot OX110 QX, UK.}
\thanks{M. Rebai was with the Department of Physics, University of Milano - Bicocca, Milano I-20126, Italy. She is now with the Istituto per la Scienza e Tecnologia dei Plasmi, CNR, Milano I-20125, Italy and with the INFN - Sezione di Milano Bicocca, Milano I-20126, Italy.}}

\maketitle

\begin{abstract}
An accurate characterization of the neutron fields at spallation sources is crucial for many applications based on neutron irradiations, such as radiation damage tests that need a precise dose estimate. In this work we present the neutron flux measurements performed with the multi-foil activation technique in the ROTAX and ChipIr beamlines of the ISIS spallation source, characterized by moderated and unmoderated spectra, respectively. We selected many different activation reactions to cover a wide energy range, from thermal to very fast neutrons up to about 100 MeV. By applying a Bayesian unfolding algorithm, we demonstrate the effectiveness of this technique in measuring the neutron flux intensity and energy spectrum with precision and accuracy.
\end{abstract}


\section{Introduction}
\label{sec:introduction}

\IEEEPARstart{S}{pallation} sources produce pulsed neutron beams by colliding high energy protons (hundreds of MeV) on heavy metal targets~\cite{Russell1995}. 
At these energies the wavelength of the proton is so short that it interacts directly with nucleus components. These interactions can be described as nucleon-nucleon collisions, rather than interactions with the whole nucleus. 
Neutrons are produced by the so-called intra-nuclear cascade, that ejects high energy nucleons up to hundreds of MeV. 
This is immediately followed by the so-called evaporation of neutrons, which are emitted with energies in the MeV range by the highly excited nuclei left by the primary collision. 
Contrary to fission and fusion, spallation is an endothermal process, which cannot be used for energy production. However, it has many advantages in producing neutrons for science.
Most importantly, the neutron fluxes can be easily controlled and pulsed via the primary beam, thus allowing to exploit the time-of-flight to detectors to retrieve neutron energy. 

Spallation sources are considered as optimum neutron sources for solid state physics and material-science investigations. For these purposes, moderators close to the targets are used to produce beams of thermal neutrons with wavelengths suitable to investigate the structure and dynamics at the atomic scale (thermal neutrons with energy $\sim25$~meV have wavelength in the order of 1~Å, the typical spacing in solid state lattices).

On the other hand, the high energy component of the neutron spectrum can be used to reproduce the effects induced by atmospheric cosmic rays in microelectronics~\cite{Cazzaniga2021}. 
Spallation sources are unique for this application, as fission or fusion cannot produce neutrons with $E>20$~MeV and up to hundreds of MeV. 
It has to be noticed that this is the same process that produces neutrons in the atmosphere, when high energy primary cosmic rays interact with nuclei at high altitudes. 
Thermal neutrons can also be relevant for microelectronics testing. Indeed, it has been found that, in commercial devices containing boron in the sensitive volume, the error rates induced by thermal neutrons can be of the same order of magnitude as those induced by fast ones~\cite{Oliveira2020}.

The study presented in this paper is driven by the fact that many applications based on neutron irradiation require a detailed knowledge of the neutron flux in terms of intensity and energy spectrum.
For example, in the irradiation tests performed to study the radiation damage on electronic devices, knowing the spectral fluence is crucial for dose calculations.
Performing an accurate measurement of the neutron flux at spallation sources is challenging, because its energy spectrum, that extends over a wide energy range up to hundreds MeV, is usually poorly known, especially in the fast region.

The various experimental techniques used for neutron flux measurement each have their pros and cons. 
Active detectors (such as gas, scintillating, or solid state detectors) are suitable for real time monitoring and time-of-flight analysis. However, time-of-flight is often not possible at high energies, because the time resolution is limited by the pulse time-width and by the path length. Moreover, active detector can suffer from saturation and pile-up (which are typical issues at pulsed sources) and a good knowledge of the detector response function is needed to determine the neutron flux.

On the other hand, the activation technique, which uses foils of different materials as passive flux monitors, does not suffer from those limitations and can provide a reliable and independent measurement of the neutron flux in a wide energy range, from the thermal region to fast one. 
Despite being a traditional method, research is needed to optimize the activation measurements and the data analysis to the specific neutron field.
In particular, the activation reactions must be chosen so as to probe the neutron spectrum in the widest possible energy range, thus allowing to minimize the dependence of the results on the guess spectrum used to unfold the flux from the activation data. 
It is also important to correctly propagate the experimental uncertainties and to assess the systematic ones, especially in the range of very fast neutrons ($E>20$~MeV), which is beyond the typical application at nuclear reactors.

In this paper, we present the neutron flux measurements performed with the multi-foil activation technique in the ROTAX and ChipIr beamlines of the ISIS spallation source of the Rutherford Appleton Laboratory (UK).
After describing the two beamlines (Sect.~\ref{sec:beamlines}), we outline the methodology that we use to determine the neutron flux intensity and characterize its energy spectrum from the activation data of multiple reactions (Sect.~\ref{sec:methodology}).
In Sect.~\ref{sec:expsetup}, we describe the experimental setup and the activation measurements performed in the ROTAX beamline.
Finally, Sect.~\ref{sec:results} includes two subsections dedicated to the discussion of the neutron flux unfolding results obtained for the ROTAX and ChipIr beamlines, respectively. 
For the latter, we present an updated analysis based on the activation data measured in a previous experimental campaign~\cite{CHIESA201814}. In particular, we unfold the neutron flux using a new guess spectrum, calculated with a Monte Carlo (MC) simulation of the spallation neutrons propagated to ChipIr.

\section{Neutron beamlines}
\label{sec:beamlines}

\begin{figure}[]
\centerline{\includegraphics[width=0.49\textwidth]{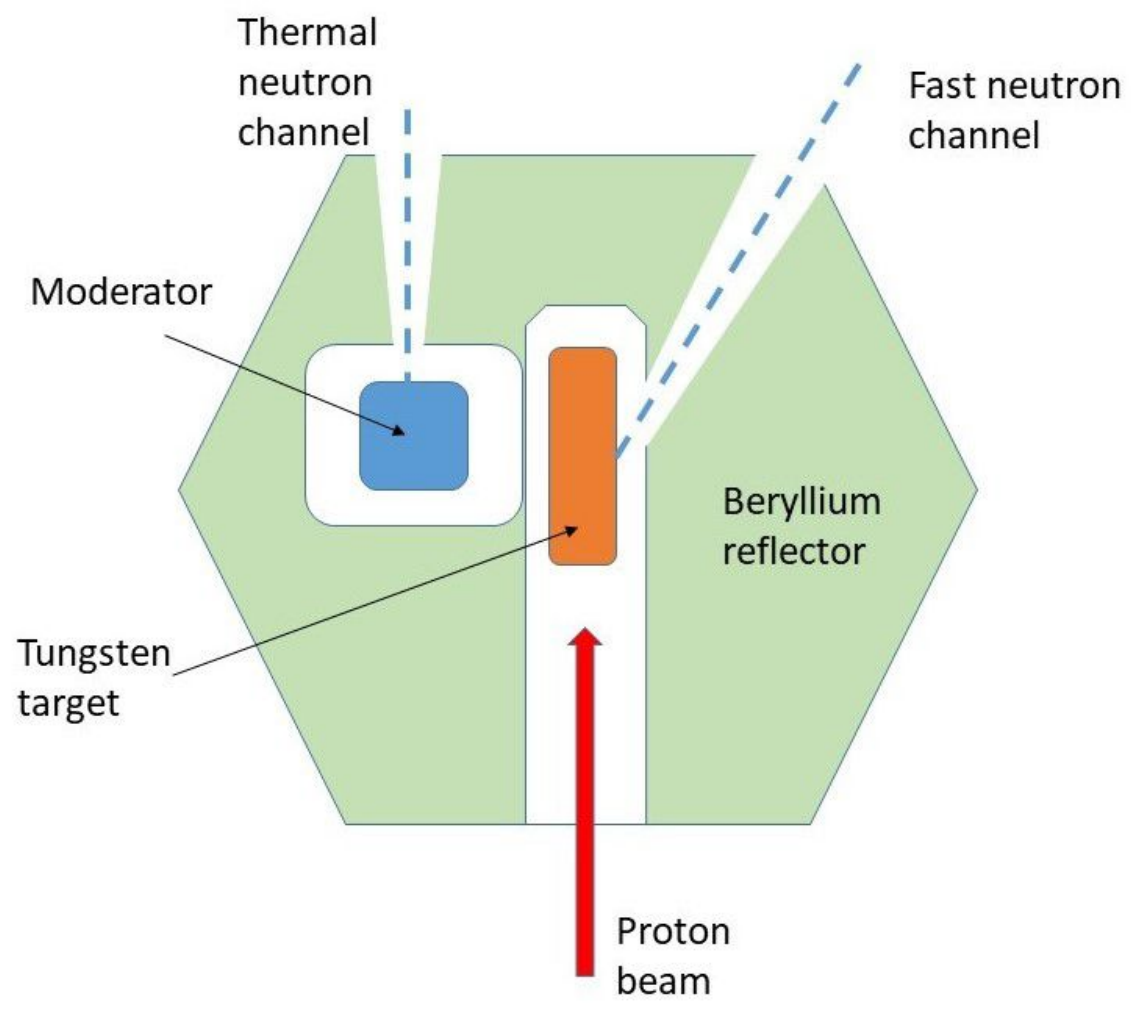}}
\caption{Simplified schematic (not to scale) of a spallation source target station featuring a fast and thermal neutron channel.}
\label{Fig:Target}
\end{figure}

The ISIS neutron spallation source of the Rutherford Appleton laboratory (UK) is based on a synchrotron running at 50~Hz and accelerating protons at 800~MeV. These impinge on tungsten targets in two target stations, named TS1 and TS2, respectively.
The proton beam current is about 160~$\mu$A on TS1 and 40~$\mu$A on TS2 
(there is 1 pulse extracted to TS2 for every 4 pulses extracted to TS1). The total yield, given by the spallation on the tungsten target, is about 20 neutrons/proton.
Neutrons are used for science on a total of 29 active beamlines.
ROTAX, now hosting the ALF instrument, is a beamline on TS1 dedicated to alignment and assessment of single crystals, mainly in support of experiments to be performed on other beamlines. It features a liquid methane moderator at 100~K. In the recent past it has been used also for Single Event Effects (SEE) testing, in particular because, compared to other beamlines at ISIS, its large blockhouse allows hosting large samples to be operated in air, with easy cable connectivity.

ChipIr is a beamline on TS2, specifically designed to deliver a beam of fast neutrons with a spectrum that mimics the atmospheric neutron field, with an acceleration factor in the order of 10$^9$ for SEE testing. ChipIr represents the first attempt to retrofit a fast neutron beamline in a target station designed for slow neutrons. 
As shown in Fig.~\ref{Fig:Target} as a schematic, while thermal neutron beamlines have a line of sight only on a moderator, ChipIr has a channel through the beryllium reflector to obtain a line of sight to the tungsten target.
The use of these beamlines allows to study the radiation damage effects using very different neutron spectra. 
This difference is also relevant with regard to the application of multi-foil activation for neutron flux measurements, because it allows us to demonstrate the flexibility and reliability of this technique in characterizing very different neutron fields.

\section{Multi-foil activation and unfolding method}
\label{sec:methodology}

The neutron activation technique is comprised of two steps. First, foils containing elements in known amounts are irradiated to produce radioisotopes with activation rates given by:
\begin{equation}
    R_j=\mathcal{N}_j\int \varphi(E)\,\sigma_j(E)\,f_j(E)\,dE
    \label{Eq:ActRate}
\end{equation}
where $\varphi(E)$ is the differential flux as a function of neutron energy, $\mathcal{N}_j$ are the number of target isotopes in the foils, $\sigma_j(E)$ are the corresponding activation cross sections, and $f_j(E)$ are corrective factors that in some cases must be introduced to keep into account the so-called \textit{self-shielding} effect and/or the presence of a Cd filter for thermal neutrons.
After irradiation, the foils are analyzed via $\gamma$-spectroscopy, typically with a Germanium detector, to identify the activated isotopes from the characteristic $\gamma$-rays and measure their activities, which are proportional to the activation rates. 
Detailed explanations of the technique are available in literature~\cite{GREENBERG2011193,AbsoluteFlux,FluxDistribution}.

The basic idea behind neutron flux unfolding is to combine the measurements of properly selected activation reactions, characterized by cross sections that allow to probe different regions of the spectrum. 
Particularly, radiative capture (n,$\gamma$) reactions have relatively high cross sections for low energy neutrons in the thermal range ($E<0.5$~eV) and are usually featured by resonances at well-defined energies in the epithermal and intermediate range.
Therefore (n,$\gamma$) reactions are excellent probes for measuring neutrons at low energies and in correspondence with the main capture resonance(s), which are distributed from 0.1~eV to a few keV depending on the target isotope.
The contribution due to fast neutrons is typically negligible for (n,$\gamma$) reactions.
On the contrary, threshold reactions in which a nucleus, following the absorption of a fast neutron, ejects other neutrons and/or charged particles (such as protons and alphas), are perfect to measure the flux in the fast region above 1~MeV. 

In order to unfold the neutron flux from the activation measurements, we apply the method described in \cite{BayesianSpectrum}.
This technique consists in solving the following system of equations with a Bayesian statistical approach:
\begin{equation}
\label{Eq:ActRateDiscrete}
\dfrac{R_j}{\mathcal{N}_j} =  \sum_{i=0}^{n} \sigma_{ij} \phi_i \quad \left(\phi_i \equiv \int_{E_i}^{E_{i+1}} \varphi(E)\,dE \right) 
\end{equation}
This system is obtained from Eq.~\ref{Eq:ActRate} by discretizing the energy spectrum into $n$ groups and its unknown variables are the \textit{group flux intensities} ($\phi_i$). The $\sigma_{ij}$ coefficients are the so called \textit{group effective cross sections}, that are calculated assuming a \textit{guess} spectral shape $\varphi(E)$ in the range of each group:
\begin{equation}
\label{Eq:XSeff}
\sigma_{ij} = \dfrac{\int_{E_i}^{E_{i+1}} \sigma_j(E)\,\varphi(E)\,f_j(E)\,dE}{\int_{E_i}^{E_{i+1}} \varphi(E)\,dE}
\end{equation}
With this unfolding method, even if the results have some degree of dependence on the \textit{intra-group} spectral shape used for $\sigma_{ij}$ calculation, there is no constraint on $\phi_i$ variables, that are left free to converge on any flux value in the positive range.

As compared with other unfolding techniques~\cite{Reginatto2010}, this method allows us to rigorously propagate the experimental uncertainties, to select the physical solutions of the problem by means of the Priors, and to analyze the correlations between the resulting $\phi_i$.
Practically, we sample the following multi-dimensional Posterior probability density function:
\begin{equation}
P\left(\phi_{i} \,|\, \dfrac{R_j}{\mathcal{N}_j}\right) = k \cdot P\left( \dfrac{R_j}{\mathcal{N}_j} \,|\, \phi_{i}\right) \cdot P\left(\phi_{i}\right)
\end{equation}
where $k$ is a normalization factor, $P ( R_j/\mathcal{N}_j \,|\, \phi_{i} )$ are Gaussian Likelihoods defined from the experimental data, and $P(\phi_{i})$ are uniform Priors in the positive range. To define the Bayesian statistical model of this problem and sample the Posterior, we use the JAGS tool that exploits Markov Chains Monte Carlo (MCMC) simulations~\cite{JAGS,JAGS_manual,Gelman}. 
By marginalizing the multi-dimensional Posterior, we extract the probability density function of each flux group $\phi_i$ and the correlations among them. The means and standard deviations of the marginalized Posteriors are finally computed to provide a multi-group neutron flux measurement, which represents the result of the unfolding procedure.

The multi-group binning of the energy spectrum is optimized by looking for a good compromise between having as many groups as possible (thus reducing the dependence on the guess spectrum) and obtaining $\phi_i$ results as uncorrelated as possible to each other. 
This is obtained by having at least a main resonance of a (n,$\gamma$) capture or the cross section peak of a threshold reaction in each group, so that there is at least one reaction induced in a non-negligible percentage by neutrons in that energy range.
On the basis of the experience gained through the application of this method to the characterization of neutron fluxes at nuclear reactors and spallation sources~\cite{BayesianSpectrum,CHIESA201814,Chiesa:2019buj}, fully satisfactory results can be obtained by measuring some tens of reactions and defining 10 energy groups or a few more.

\section{Experimental setup and measurements}
\label{sec:expsetup}

\begin{table}[]
\caption{List of foils irradiated in ROTAX}
\begin{tabular}{lccc}
\hline
Foil & Mass (g) & Thickness (mm) & $t_{\text{irr}}$ (h)\\
\hline
NaCl	&	0.4647	&	1.905	&	1.002	\\
Al	&	0.2548	&	0.127	&	0.249	\\
Sc	&	0.0502	&	0.127	&	1.871	\\
V	&	0.0439	&	0.051	&	0.254	\\
Mn-Cu (Mn 83.1\%)	&	0.0495	&	0.127	&	0.618	\\
Co	&	0.0662	&	0.051	&	4.404	\\
ZnSe	&	0.3908	&	2.0	&	13.98	\\
Zr	&	0.1114	&	0.127	&	2.549	\\
Mo	&	0.0923	&	0.076	&	0.905	\\
In	&	0.1296	&	0.127	&	3.179	\\
Lu-Al (Lu 5.1\%)	&	0.0336	&	0.102	&	15.16	\\
W	&	0.3247	&	0.127	&	0.563	\\
Au	&	0.1206	&	0.051	&	0.600	\\
Au-Al (Au 0.134\%)	&	0.0445	&	0.127	&	12.86	\\
\hline
NaCl (Cd cover)	&	0.4717	&	1.905	&	0.841	\\
Al (Cd cover)	&	0.2328	&	0.127	&	0.333	\\
V (Cd cover)	&	0.0439	&	0.051	&	0.267	\\
Ni (Cd cover)	&	0.2816	&	0.254	&	3.890	\\
Zn (Cd cover)	&	0.2320	&	0.254	&	13.90	\\
ZnSe (Cd cover)	&	0.2159	&	2.000	&	0.950	\\
Au (Cd cover)	&	0.1177	&	0.051	&	0.472	\\
\hline
Mg	&	0.0309	&	0.127	&	15.16	\\
Al	&	0.2631	&	1.0	&	5.727	\\
Fe	&	0.8325	&	1.0	&	1.871	\\
Co	&	0.4761	&	0.5	&	13.90	\\
Ni	&	0.9700	&	1.0	&	15.16	\\
In	&	0.1255	&	0.127	&	13.90	\\
Bi	&	1.0538	&	1.0	&	13.98	\\
\hline
\multicolumn{4}{p{0.45\textwidth}}{The weight and the thickness of each sample is reported together with the irradiation time, $t_{\text{irr}}$. In the first (second) section we list the foils irradiated in the front position without (with) the cadmium cover. In the third section we list the samples irradiated in the rear positions of the holder.}\\
\end{tabular}
\label{tab:ListFoils}
\end{table}

\begin{figure}[]
\centerline{\includegraphics[width=0.45\textwidth]{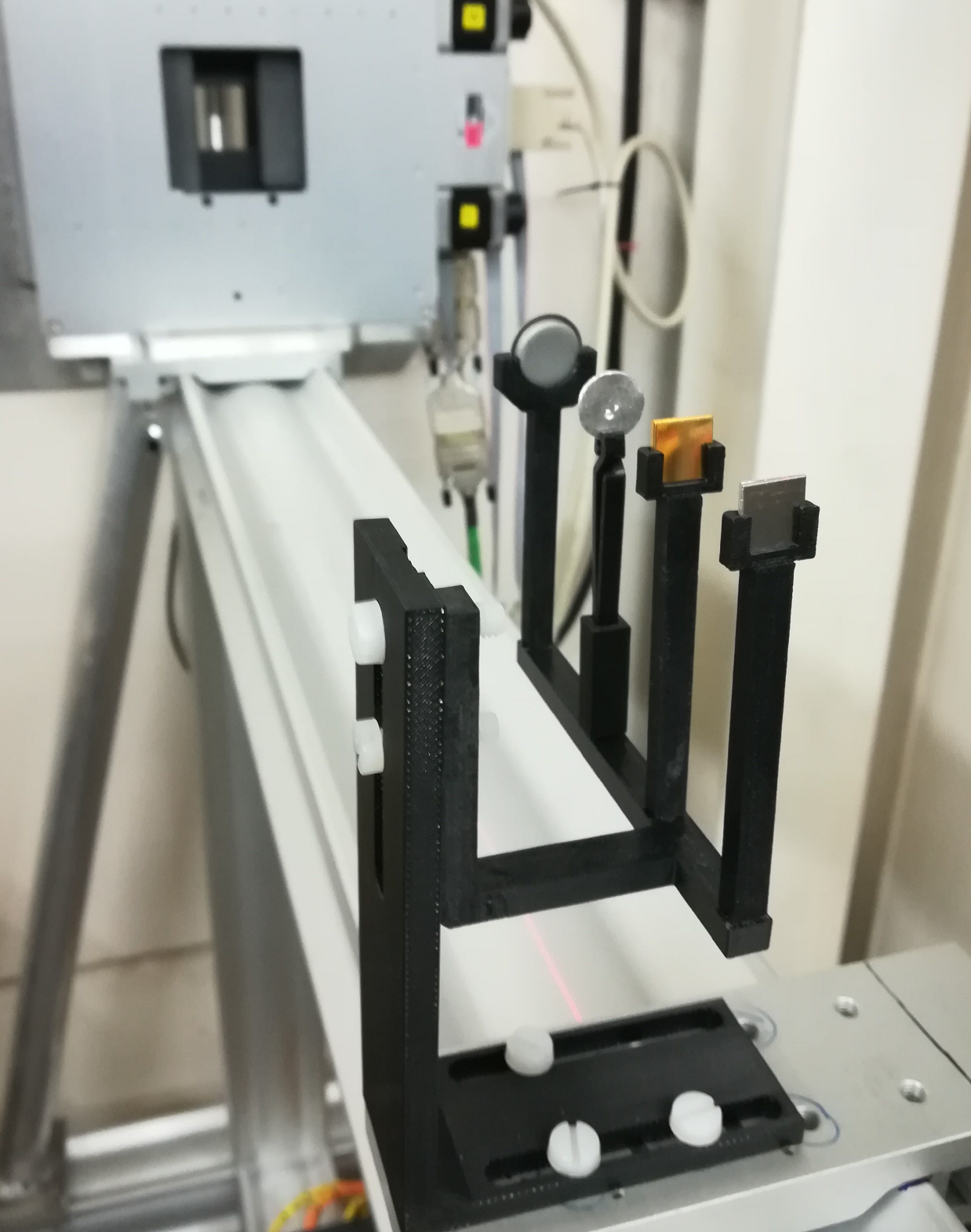}}
\caption{Experimental setup used for multi-foil activation in the ROTAX beamline. The black plastic sample holder in the foreground allows to irradiate four foils at a time and was positioned $\sim$90 cm apart from the beam port visible in the background.}
\label{Fig:Irradiation}
\end{figure}

To measure the flux in the ROTAX and ChipIr beamlines, we performed two irradiation campaigns of about 3 days each. The foils were mounted on ad hoc plastic holders designed to minimize flux perturbation. 
The activation measurements at ChipIr have been performed in November 2016 (with the ISIS accelarator working at 700~MeV) and are fully documented in \cite{CHIESA201814}.
In June 2018, we performed a second activation campaign in the ROTAX beamline, with the proton beam accelerated at 800~MeV. 
As compared with the previous activation campaign, we have made some improvements in the experimental setup, that was designed to irradiate up to four samples at a time, as shown in Fig.~\ref{Fig:Irradiation}.
We used the front position for measuring (n,$\gamma$) reactions, and we exploited the rear ones for threshold reactions since fast neutrons are negligibly absorbed or scattered by the interposed foils. 
Moreover, in order to enhance the experimental sensitivity in measuring the neutron flux in the intermediate energy range, we irradiated 7 foils inside Cd covers that completely absorb thermal neutrons, thus allowing for the (n,$\gamma$) reactions to be activated only by neutrons at resonances. 
The list of 28 foils irradiated in the ROTAX beamline is reported in \textsc{Table}~\ref{tab:ListFoils}, with their weights, thicknesses and irradiation times.
We used samples small enough to fit into the $2~\text{cm}\times2~\text{cm}$ beam area in which the neutron flux is uniform~\cite{ROTAXbeamarea}.
In particular, all the foils are round shaped with a diameter of 1.27~cm, except for the Al, Fe, Co, Ni, and Bi ones (used for threshold reactions), which are squares $1~\text{cm}\times1~\text{cm}$.
Their purity and elemental compositions are certified by the manufacturers~\cite{Shieldwerx, GoodFellow}.

In order to calculate the net irradiation times keeping into account possible beam interruptions, we monitored the neutron flux with a diamond detector installed behind the activation foils. In this way, we could also determine with precision the opening and closing times of the beamline shutter.

\begin{figure}[]
\subfigure{\includegraphics[width=0.24\textwidth]{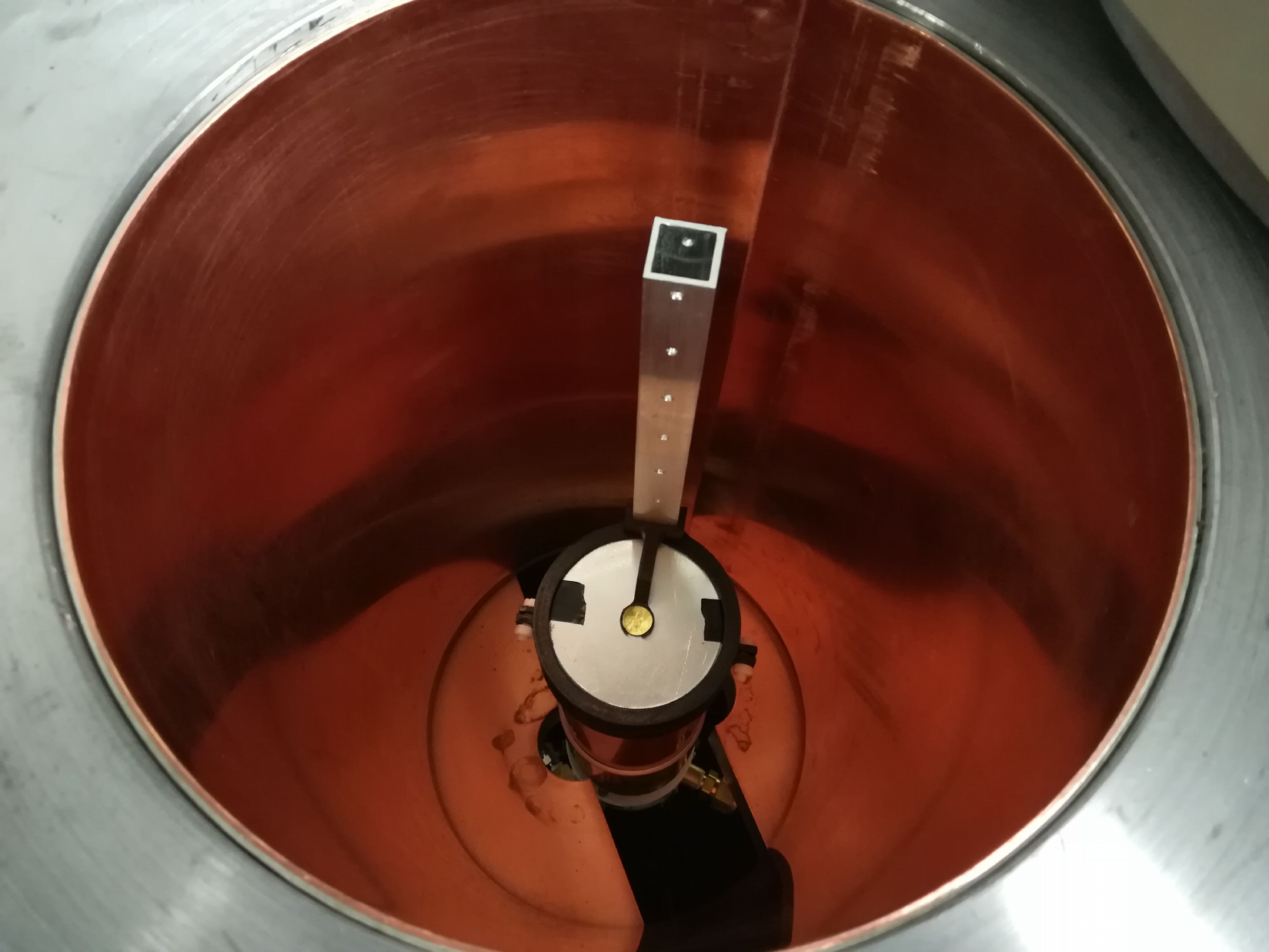}}
\subfigure{\includegraphics[width=0.24\textwidth]{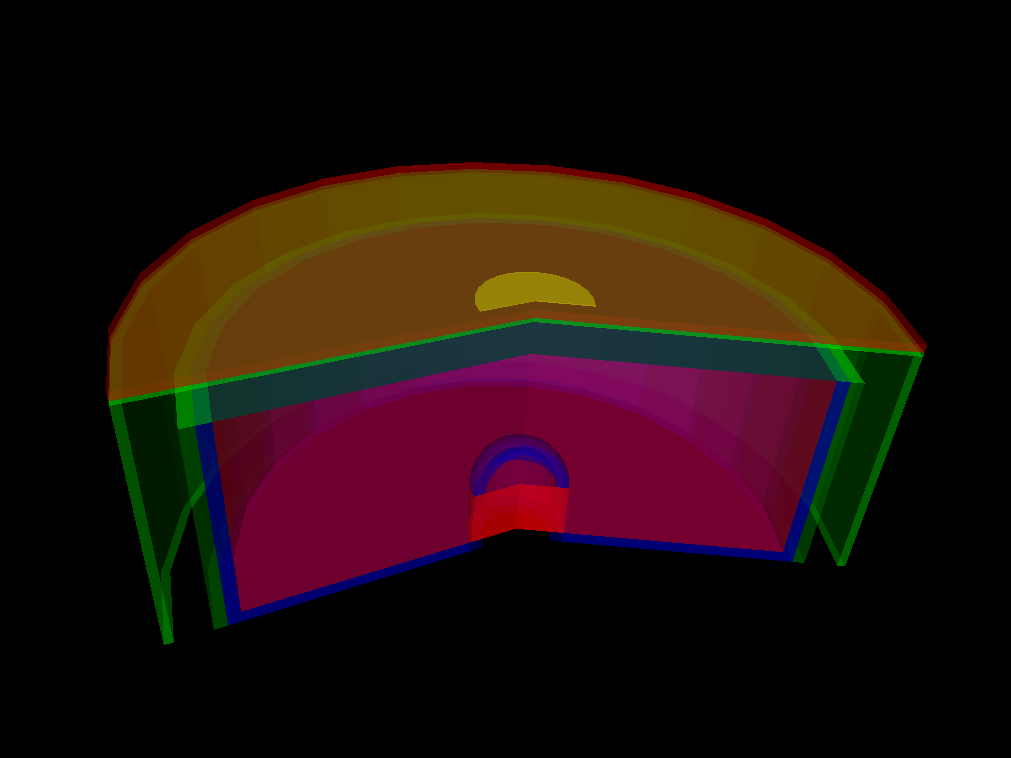}}
\caption{The HPGe detector, equipped with the sample holder, used for the $\gamma$-spectroscopy measurements of the foils activated in ROTAX (left), and its reconstruction in Geant4-based MC simulations (right).}
\label{Fig:HPGe}
\end{figure}

After the irradiations, we performed $\gamma$-spectroscopy measurements with a High Purity Germanium (HPGe) detector (Fig.~\ref{Fig:HPGe}, left). 
This detector, belonging to the GEM series by ORTEC, is coaxial type with beryllium endcap and 20\% nominal efficiency at 1.33~MeV. 
We added a 1~mm thick aluminum layer above the beryllium endcap to shield the detector from the low energy X-rays, that otherwise could sum up with the characteristic $\gamma$s (thus complicating the analysis). 
In order to ensure a precise sample positioning, we manufactured a holder that allowed us to place the sample at different distances (depending on its activity) along the central axis of the detector.
The $\gamma$-spectroscopy measurements lasted for about a month and most of the foils were measured more than once just after the end of their irradiation and/or after some cooling time to be sensitive to both short-lived and long-lived isotopes.

\section{Data analysis and results}
\label{sec:results}

\subsection{ROTAX beamline}

\begin{table}[b!]
\caption{Activation rates per unit mass of radiative capture reactions in the ROTAX beamline}
\begin{tabular}{lcc}
\hline
Reaction & $R/m$ (s$^{-1}$g$^{-1}$) & $R/m$ (s$^{-1}$g$^{-1}$)\\
         & without Cd cover             & with Cd cover           \\
\hline
$^{23}$Na (n,$\gamma$)	&	$(2.53\pm0.08)\times10^{4}$	&	$(5.05\pm0.17)\times10^{3}$	\\
$^{27}$Al (n,$\gamma$)	&	$(7.80\pm0.38)\times10^{3}$	&	$(1.36\pm0.14)\times10^{3}$	\\
$^{37}$Cl (n,$\gamma$)	&	$(3.14\pm0.10)\times10^{3}$	&	$(5.11\pm0.36)\times10^{2}$	\\
$^{45}$Sc (n,$\gamma$)	&	$(6.09\pm0.18)\times10^{5}$	& \\
$^{51}$V (n,$\gamma$)	&	$(1.03\pm0.04)\times10^{5}$	&	$(1.74\pm0.06)\times10^{4}$	\\
$^{55}$Mn (n,$\gamma$)	&	$(3.10\pm0.10)\times10^{5}$	& \\
$^{59}$Co (n,$\gamma$)	&	$(1.00\pm0.03)\times10^{6}$	& \\
$^{64}$Ni (n,$\gamma$)	&                               &   $(4.39\pm0.30)\times10^{1}$	\\
$^{64}$Zn (n,$\gamma$)	&	$(9.53\pm0.36)\times10^{3}$	&	$(4.91\pm0.17)\times10^{3}$ \\
$^{74}$Se (n,$\gamma$)	&	$(2.76\pm0.08)\times10^{4}$	&   $(2.36\pm0.08)\times10^{4}$ \\
$^{80}$Se (n,$\gamma$)	&	$(6.88\pm0.30)\times10^{3}$ &	$(4.01\pm0.24)\times10^{3}$ \\
$^{82}$Se (n,$\gamma$)	&                               &	$(1.18\pm0.32)\times10^{2}$	\\
$^{96}$Zr (n,$\gamma$)	&	$(7.02\pm0.23)\times10^{2}$	& \\
$^{100}$Mo (n,$\gamma$)	&	$(1.89\pm0.06)\times10^{3}$	& \\
$^{113}$In (n,$\gamma$)	&	$(5.07\pm0.36)\times10^{4}$	& \\
$^{115}$In (n,$\gamma$)	&	$(3.92\pm0.19)\times10^{6}$	& \\
$^{176}$Lu (n,$\gamma$)	&	$(6.14\pm0.37)\times10^{5}$	& \\
$^{186}$W (n,$\gamma$)	&	$(1.88\pm0.06)\times10^{5}$	& \\
$^{197}$Au (n,$\gamma$)	&	$(1.58\pm0.05)\times10^{6}$	&	$(1.14\pm0.04)\times10^{6}$ \\
$^{197}$Au (n,$\gamma$) in Au-Al	&	$(3.32\pm0.12)\times10^{6}$	& \\
\hline
\multicolumn{3}{p{0.45\textwidth}}{The activation rates per unit mass of the target element in the foil are reported in the second and third column for irradiations performed with and without Cd cover, respectively. When a (n,$\gamma$) reaction activates isotopes in both the ground and metastable states, the experimental data are properly scaled for the isomeric branching ratios to get the total activation rate. The reported uncertainties include the statistical component, the isomeric branching ratio uncertainty (if used to scale the data), and a 3\% systematic error associated to the efficiency reconstruction in $\gamma$-spectroscopy measurements.}\\
\end{tabular}
\label{tab:ListReactions}
\end{table}

In order to analyze the $\gamma$ spectra acquired in the experimental campaign at ROTAX and determine the activation rates $R_j$, we apply the same procedure used for the measurements previously performed at ChipIr and documented in detail in~\cite{CHIESA201814}.
We recall here the two most important aspects that allow us to obtain precise and accurate measurements of the activation rates.

First, since the isotope activity must be determined on absolute scale from the counts recorded at the full-energy $\gamma$ peaks, we need an accurate efficiency evaluation for each $\gamma$-line in the spectrum. For this purpose, we exploit a Geant4-based MC software~\cite{GEANT4}, that allows us to flexibly implement the geometry and the materials of the experimental setup (detector and sample) and that provides in the output a simulated spectrum corresponding to a certain number of decays of any radioisotope of interest. In Fig.~\ref{Fig:HPGe} (right), we show the MC model of the detector, implemented according to the geometry provided by the manufacturer. 
We validated the MC simulations through benchmark measurements with certified-activity multi-$\gamma$ sources ($^{60}$Co, $^{137}$Cs, and $^{241}$Am), obtaining an average discrepancy of 3\% on the efficiency reconstruction of the different $\gamma$-lines.

Second, when we analyze the activated foils and different peaks produced by the same isotope are observed in the spectrum, we combine and cross check the activity measurements obtained from each of them to improve the accuracy and precision in activation rate assessment. Similarly, we combine and cross check the experimental results from repeated measurements of the same foil at different cooling times.

\begin{table}[]
\caption{Activation rates per unit mass of threshold reactions in the ROTAX beamline}
\begin{tabular}{lcc}
\hline
Reaction  & Threshold (MeV) & $R/m$ (s$^{-1}$g$^{-1}$) \\
\hline		
Mg (n,*) $^{24}$Na	        & 6 &	$(9.09\pm0.72)\times10^{1}$	\\
Al (n,$\alpha$) $^{24}$Na	& 6 &	$(6.47\pm0.27)\times10^{1}$	\\
Al (n,p) $^{27}$Mg	        & 3 &	$(1.05\pm0.07)\times10^{2}$	\\
Fe (n,*) $^{56}$Mn	        & 5 &	$(3.19\pm0.17)\times10^{1}$	\\
Co (n,2n) $^{58}$Co	        & 11 &	$(1.37\pm0.10)\times10^{2}$ \\
Co (n,3n) $^{57}$Co		    & 20 &	$(3.54\pm0.51)\times10^{1}$ \\
Co (n,$\alpha$) $^{56}$Mn	& 6  &	$(1.06\pm0.07)\times10^{1}$	\\
Ni (n,*) $^{57}$Ni	        & 13 &	$(1.51\pm0.09)\times10^{1}$	\\
Ni (n,*) $^{58}$Co		    & 0.9 &	$(5.08\pm0.16)\times10^{2}$ \\
Ni (n,*) $^{57}$Co	        & 10  &	$(1.39\pm0.05)\times10^{2}$	 \\
Ni (n,*) $^{56}$Co		    & 19  &	$(2.54\pm0.20)\times10^{1}$ \\
In (n,n') $^{115m}$In       & 0.4 &	$(7.89\pm0.27)\times10^{2}$ \\
Bi (n,4n) $^{206}$Bi	    & 24  &	$32.0\pm1.1$ \\
Bi (n,5n) $^{205}$Bi        & 32  &	$24.5\pm2.1$ \\
Bi (n,6n) $^{204}$Bi	    & 40  &	$14.5\pm0.7$ \\
Bi (n,7n) $^{203}$Bi		& 48  &	$9.1\pm0.6$  \\
Bi (n,8n) $^{202}$Bi		& 56  &	$6.0\pm0.9$  \\
\hline
\multicolumn{3}{p{0.45\textwidth}}{The activation rates are expressed per unit mass of the target element in the foil. The reported uncertainties include the statistical component and a 3\% systematic error associated to the MC efficiency reconstruction.}\\
\end{tabular}
\label{tab:ListThresholdReact}
\end{table}

In \textsc{Tables} \ref{tab:ListReactions} and \ref{tab:ListThresholdReact}, we report the activation rates per unit mass of the target element ($R/m$) for the (n,$\gamma$) and the threshold reactions measured in the ROTAX beamline.
When metastable states of the compound nucleus can be activated following a neutron capture, we keep into account the corresponding isomeric branching ratios, taken from the Atlas of Neutron Capture Cross Sections by IAEA~\cite{ngatlas}, to get the total (n,$\gamma$) activation rates.
We calculate the uncertainties on $R/m$ by combining the statistical Poisson fluctuations of the peak counts, the isomeric branching ratio uncertainty (if used to scale the data), and a 3\% systematic error related to the MC efficiency reconstruction.

\begin{figure}[]
\centerline{\includegraphics[width=0.45\textwidth]{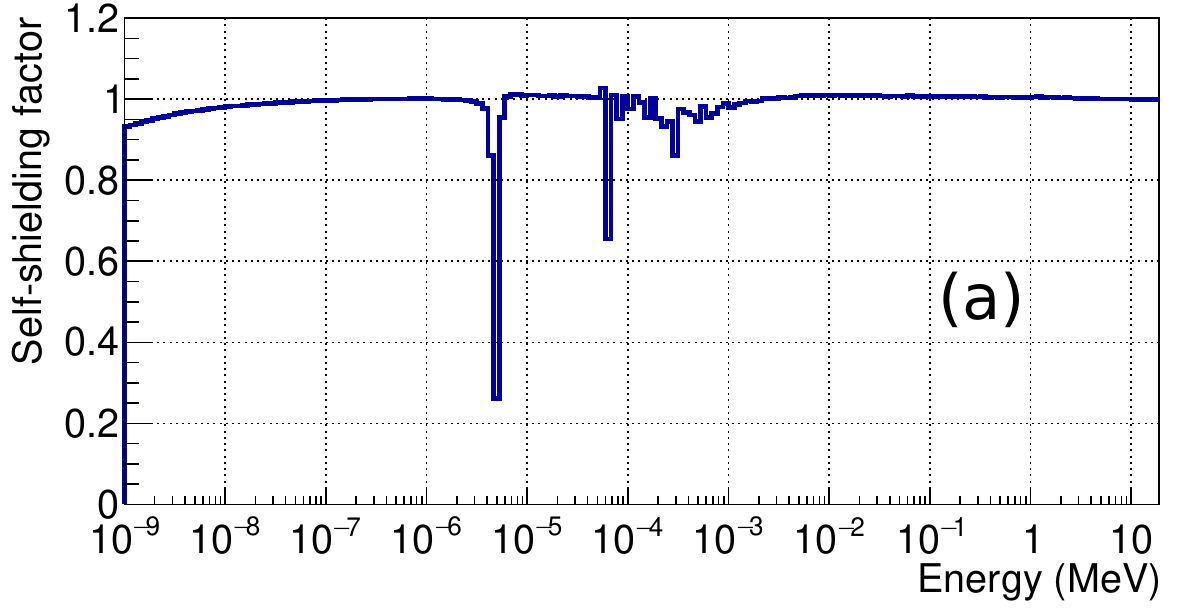}}
\centerline{\includegraphics[width=0.45\textwidth]{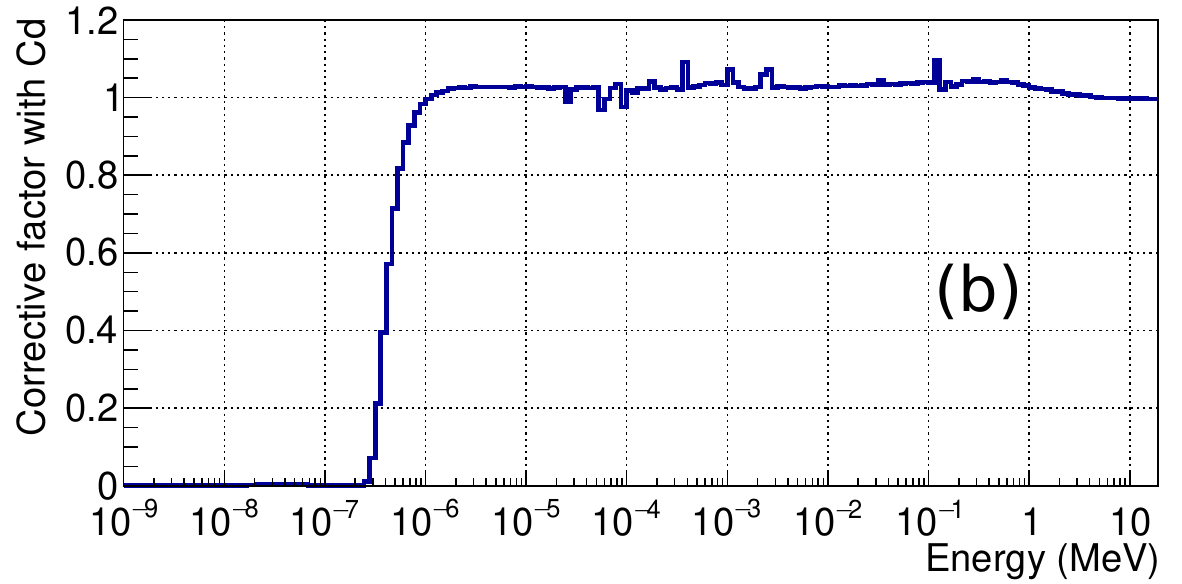}}
\caption{(a) Self-shielding factor of the Au foil as a function of neutron energy for the $^{197}$Au (n,$\gamma$) reaction, calculated as the ratio of the activation rates obtained from MCNP simulations of a real-density and a zero-density foil. The self-shielding effect is clearly visible in correspondence of the main resonances of the $^{197}$Au (n,$\gamma$) cross section.
(b) Corrective factor for the NaCl foil irradiated inside the 0.5~mm thick Cd cover, obtained from MCNP simulations that account for both absorption and scattering effects. Thermal neutrons below Cd cut-off ($\sim$0.5~eV) are completely absorbed, whereas the flux of resonance and fast neutrons is negligibly modified.}
\label{Fig:SelfShield}
\end{figure}

In total, we measured 18 (n,$\gamma$) reactions without Cd cover, 10 reactions with Cd cover and 17 threshold reactions. It is worth to point out that we measured high-threshold (n,$x$n) reactions on Bi target, that allows us to probe the flux in the region of very fast neutrons with $E>20$~MeV.

In order to unfold the neutron flux in the ROTAX beamline, we define a multi-group binning (see second column of \textsc{Table}~\ref{tab:RotaxUnfolding}) with the criteria outlined in Sect.~\ref{sec:methodology}. 
Then, we calculate the group effective cross sections $\sigma_{ij}$ using the $\sigma(E)$ data published in the ENDF/B-VII.1~\cite{ENDF} and TENDL~\cite{TENDL} libraries for (n,$\gamma$) captures and threshold reactions, respectively. Since the TENDL database provides cross section evaluations up to 200~MeV, we perform a linear extrapolation in the last group that extends to 800~MeV\footnote{Since at $E>100$~MeV the cross sections of the observed reactions are relatively low and the neutron flux has a decreasing trend, we can state that this arbitrary assumption does not significantly affect the final results.}.

The guess spectrum $\varphi(E)$ that we use to compute $\sigma_{ij}$ is obtained from MCNP6~\cite{mcnp6} simulations aimed at reproducing the complexity of neutron production in the spallation target, the moderation process in the liquid methane, and the neutron transport in the direction of the ROTAX beamline. 

In order to determine the corrective factors $f_j(E)$ for self-shielding and Cd cover effects, we run MCNP simulations in which we exploit the FM card to get binned spectra of the activation rates as a function of neutron energy and we calculate, bin-by-bin, the ratio between these rates to those calculated with the unperturbed neutron flux. 
In Fig.~\ref{Fig:SelfShield}, we show two plots of the $f_j(E)$ factors that we obtained from simulations to keep into account the self-shielding effect in a pure gold foil and the Cd cover filtering action on thermal neutrons.

Finally, we run the unfolding algorithm presented in Sect.~\ref{sec:methodology} to get the multi-group neutron flux in the ROTAX beamline. 
Since the last group comprising the neutrons with $E>100$~MeV cannot be determined from the experimental data of the measured reactions, we constrained it to connect with continuity with the previous one.
Moreover, to take into account the uncertainties associated with cross sections and propagate them to the flux results, we add Gaussian likelihoods in the Bayesian model to let the $\sigma_{ij}$ coefficients free to float around their central values. In particular, we take the uncertainties of thermal neutron capture cross sections and resonance integrals from the BNL-98403-2012-JA Report~\cite{PRITYCHENKO20123120}, based on the Low Fidelity Covariance Project~\cite{LITTLE20082828}, whereas we use the uncertainties provided by the TENDL library for threshold reactions\footnote{When the uncertainty is not available, as in the case of some (n,$x$n) reactions, we set it to 30\%, which corresponds to the relative uncertainty quoted in the TENDL database for similar reactions.}.

\begin{figure}[]
\centerline{\includegraphics[width=0.4\textwidth]{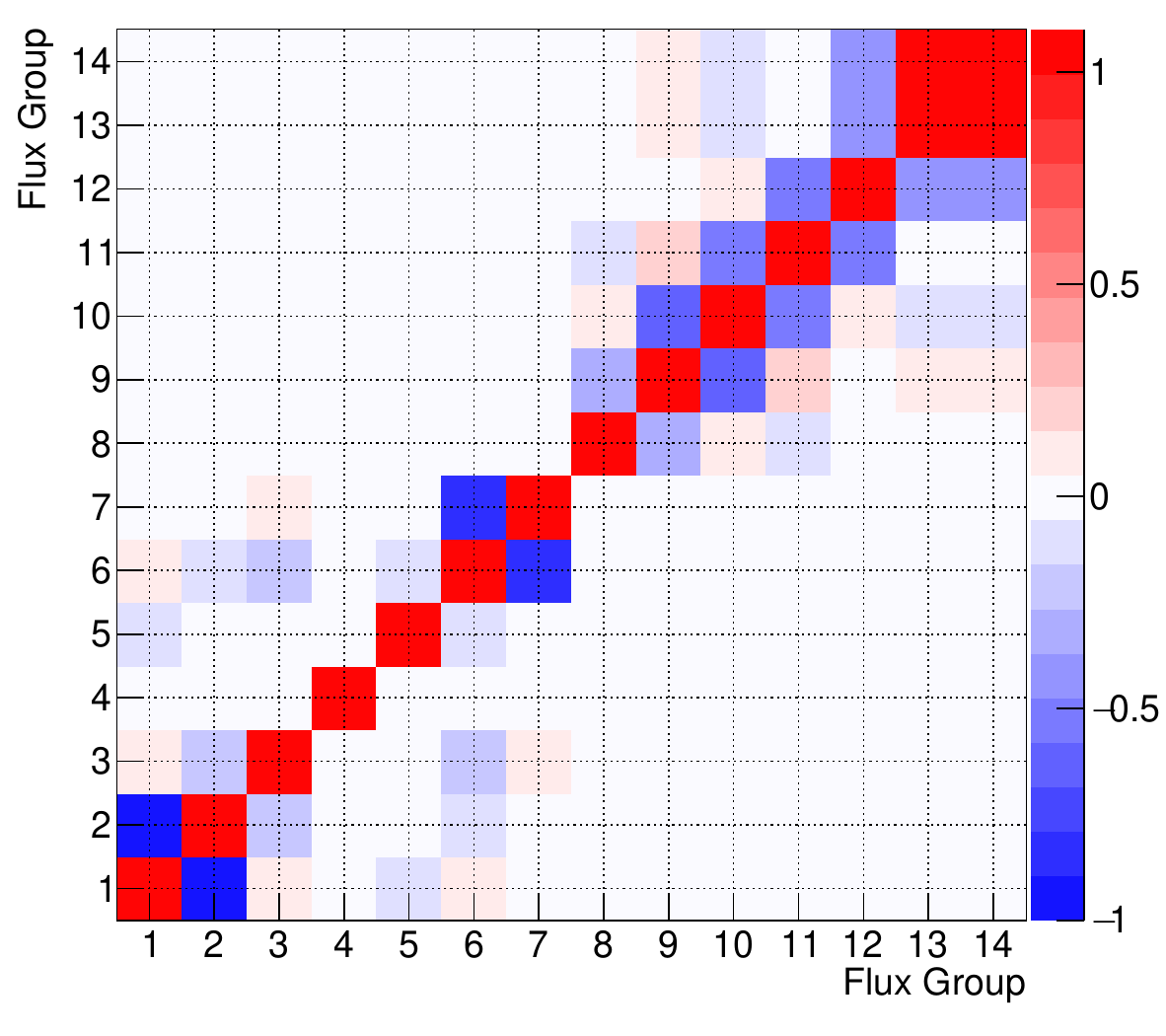}}
\caption{Correlation matrix of the unfolded flux group intensities (ROTAX beamline).}
\label{Fig:RotaxCorrelations}
\end{figure}

In \textsc{Table}~\ref{tab:RotaxUnfolding} we present the multi-group neutron flux intensities ($\phi_i$) resulting from the unfolding, and in Fig.~\ref{Fig:RotaxCorrelations} the correlations among them. Apart from a few couples of adjacent groups which are clearly anti-correlated (meaning that their sum is determined more precisely than each of them taken separately), most of the $\phi_i$ are negligibly correlated to each other, thus demonstrating the effectiveness of the experimental technique in measuring the neutron flux in the different energy ranges.

\begin{figure}[]
\centerline{\includegraphics[width=0.45\textwidth]{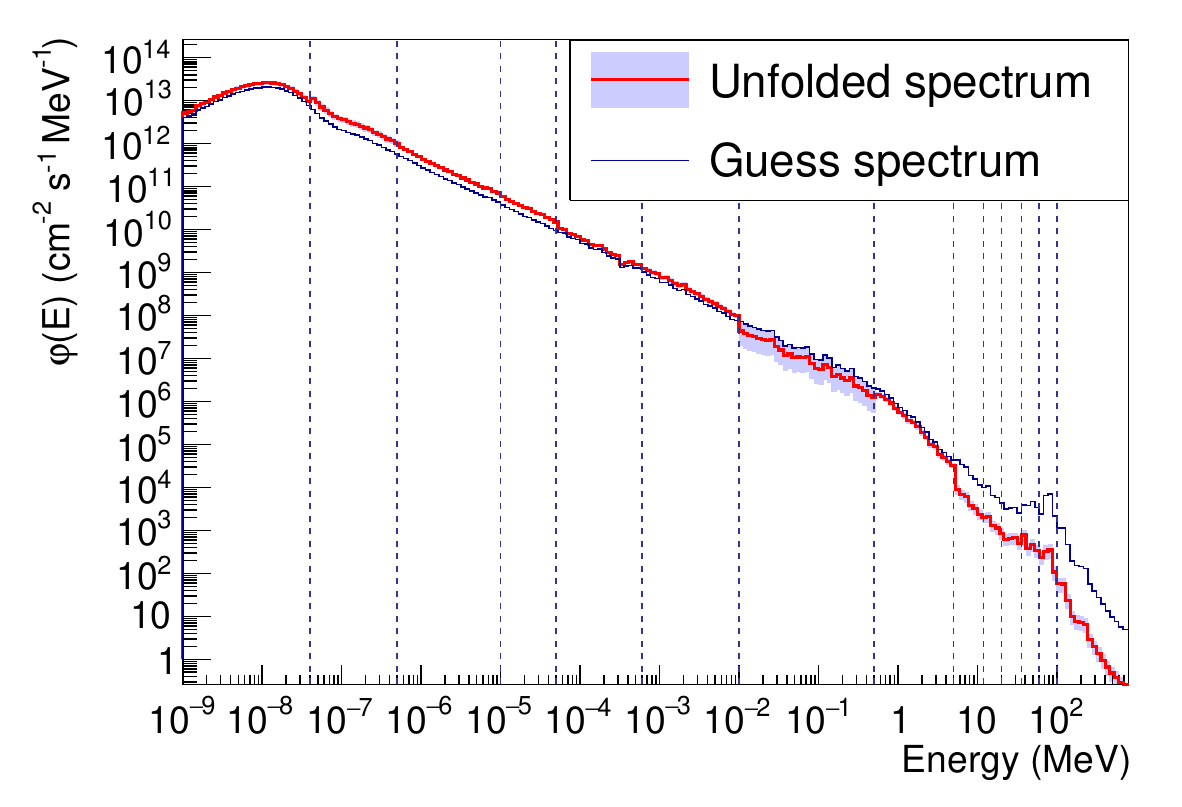}}
\caption{Unfolded neutron flux spectrum in the ROTAX beamline. The multi-group binning is shown by the dotted vertical lines. The red histogram is obtained by normalizing the guess spectrum groups to the multi-group flux intensities reported in \textsc{Table}~\ref{tab:RotaxUnfolding}. The light blue shaded area represents the uncertainty of the flux intensity in each group.
The blue histogram is the guess spectrum used for the unfolding, normalized to the same integral of the unfolded flux.}
\label{Fig:RotaxUnfolding}
\end{figure}

In Fig.~\ref{Fig:RotaxUnfolding}, we provide a graphical representation of the unfolding result by normalizing the guess spectrum of each group to the corresponding $\phi_i$ intensity. In the same plot, we also show the guess spectrum normalized to the total flux.
It is worth noting that most of the group fluxes result to be connected with continuity (compatibly with the statistical uncertainty represented in light blue) even if there is no continuity constraint in the unfolding algorithm.
Particularly, we observe a very good agreement between the experimental measurement and the simulated guess spectrum in the thermal and intermediate region of the spectrum. This is an important result that represents a validation of the experimental technique and, at the same time, a benchmark for MC simulations.
On the other hand, in the fast region above 5~MeV, the measured flux is significantly lower with respect to the guess one. This difference can be traced back to the fact that the guess spectrum is obtained from a simulation that, to reduce the computing cost, tallies the neutron flux at about 3.7 m from the moderator, whereas the ROTAX beamline is about 15 m long. Nevertheless, thanks to the variety of the measured threshold reactions, we can unfold the fast flux with 6 independent groups from 0.5 MeV to 100 MeV, reconstructing its spectrum with good detail, and determining its intensity with accuracy.

\begin{table}[]
\caption{Results of the neutron flux unfolding in the ROTAX beamline}
\begin{tabular}{c|ll|ll}
\hline
Group & \multicolumn{2}{|c|}{Energy range (MeV)} & \multicolumn{2}{|c|}{Neutron Flux (cm$^{-2}$s$^{-1}$)} \\
\hline
1 &  $10^{-9}$          &   $4\times10^{-8}$    & $(7.05\pm0.56)\times10^{5}$ &   (8.0\%)  \\ 
2 &  $4\times10^{-8}$   &   $5\times10^{-7}$    & $(1.10\pm0.16)\times10^{6}$ &   (15\%) \\ 
3 &  $5\times10^{-7}$   &   $10^{-5}$           & $(1.56\pm0.04)\times10^{6}$ &   (2.6\%) \\ 
4 &  $10^{-5}$          &   $5\times10^{-5}$    & $(1.09\pm0.04)\times10^{6}$ &   (3.5\%) \\ 
5 &  $5\times10^{-5}$   &   $6\times10^{-4}$    & $(1.66\pm0.06)\times10^{6}$ &   (3.7\%)  \\ 
6 &  $6\times10^{-4}$   &   $10^{-2}$           & $(2.54\pm0.26)\times10^{6}$ &   (10\%) \\ 
7 &  $10^{-2}$          &   0.5                 & $(2.5\pm1.3)\times10^{6}$ &   (55\%) \\ 
8 &  0.5                &   5                   & $(1.12\pm0.14)\times10^{6}$ &   (13\%) \\ 
9 &  5                  &   12                  & $(3.2\pm0.8)\times10^{4}$ &   (24\%) \\ 
10 &  12                &   20                  & $(1.2\pm0.3)\times10^{4}$ &   (28\%) \\ 
11 &  20                &   36                  & $(1.0\pm0.3)\times10^{4}$ &   (28\%) \\ 
12 &  36                &   60                  & $(9.5\pm2.8)\times10^{3}$ &   (30\%) \\ 
13 &  60                &   100                 & $(9.0\pm3.2)\times10^{3}$ &   (35\%) \\ 
14 &  100               &   800                 & $(3.2\pm1.1)\times10^{3}$ &   (35\%) \\ 
\hline        
\multicolumn{2}{c}{}&  Total     & $(1.23\pm0.11)\times10^{7}$ &   (9.3\%)  \\
\hline
\multicolumn{5}{p{0.45\textwidth}}{The uncertainties are statistical ones only, obtained by propagating the activation rates and the cross section uncertainties. In the last column we report the corresponding relative uncertainties.}\\
\end{tabular}
\label{tab:RotaxUnfolding}
\end{table}

\subsection{ChipIr beamline}

In this subsection we present the results of a new unfolding analysis of the neutron flux in the ChipIr beamline, using the activation data published in \cite{CHIESA201814}.
The main novelty is that we now have a MCNP6 simulation that gives us a MC-based guess spectrum in the range at $E>0.1$~MeV. 
At lower energies, we compute the effective cross sections with the same spectral shape used for the previous unfolding \cite{CHIESA201814}, i.e. constant flux in the first group up to 0.04~eV and $\varphi(E) \propto E^{-1}$ from 0.04~eV to 0.1~MeV.

\begin{figure}[]
\centerline{\includegraphics[width=0.4\textwidth]{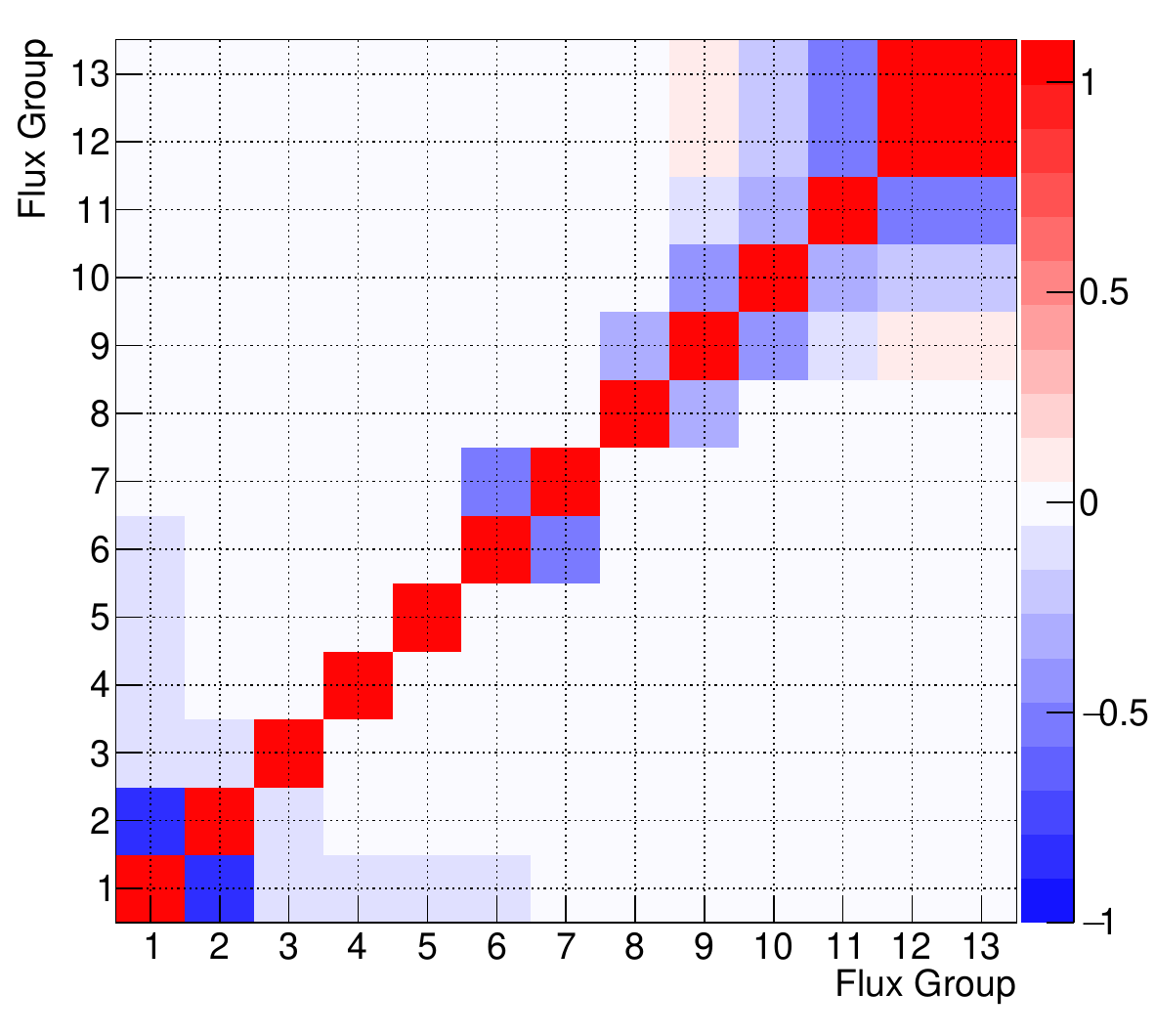}}
\caption{Correlation matrix of the unfolded flux group intensities (ChipIr beamline).}
\label{Fig:ChipIrCorrelations}
\end{figure}

We report the $\phi_i$ unfolding results in \textsc{Table}~\ref{tab:ChipIrUnfolding} and their correlations in Fig.~\ref{Fig:ChipIrCorrelations}.
When compared with the previous unfolding version, the new $\phi_i$ results are all within the ranges of systematic uncertainties investigated in \cite{CHIESA201814}.
The quality of the results in terms of precision and structure of anti-correlations is similar to that obtained for ROTAX, even if these beamlines are characterized by very different spectra.
This result highlights the versatility of the multi-foil activation technique in measuring different neutron fields.

\begin{figure}[]
\centerline{\includegraphics[width=0.45\textwidth]{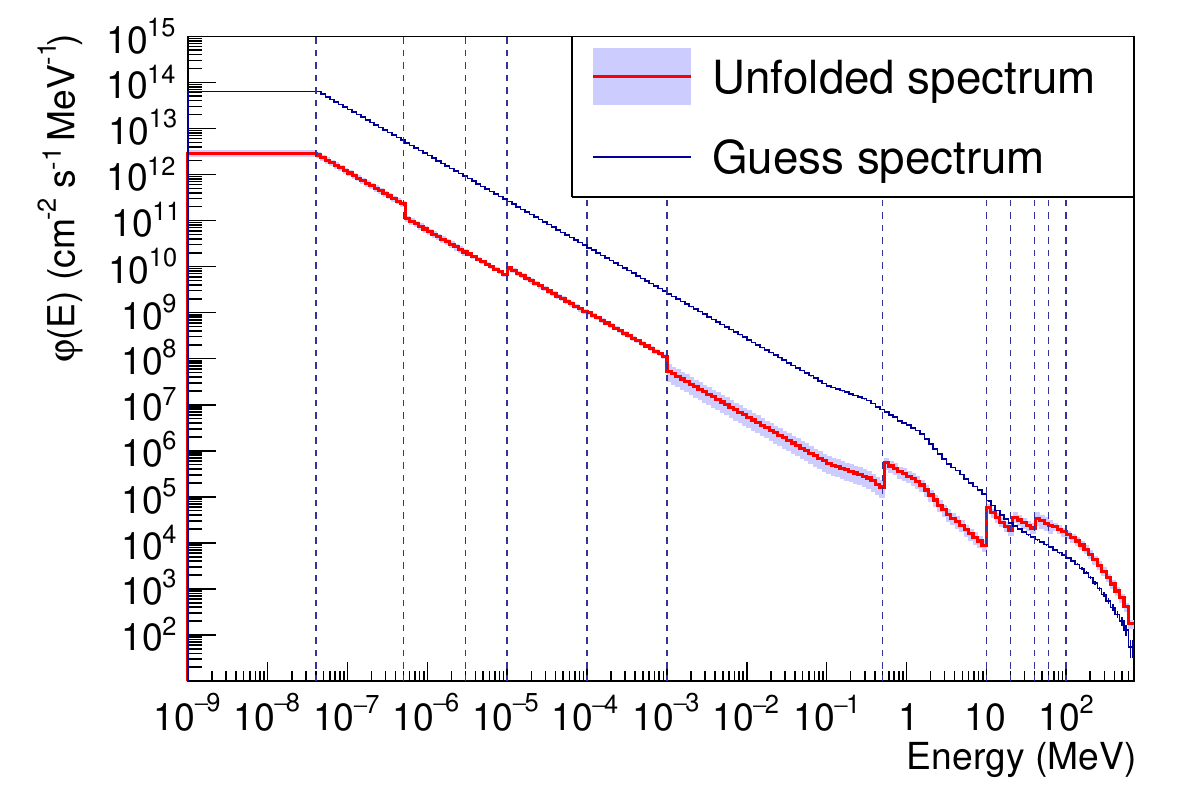}}
\caption{Unfolded neutron flux spectrum in the ChipIr beamline. The multi-group binning is shown by the dotted vertical lines. The red histogram is obtained by normalizing the guess spectrum groups to the multi-group flux intensities reported in \textsc{Table}~\ref{tab:ChipIrUnfolding}. The light blue shaded area represents the uncertainty of the flux intensity in each group.
The blue histogram is the guess spectrum used for the unfolding, normalized in the fast region as results from the MCNP6 simulation.}
\label{Fig:ChipIrUnfolding}
\end{figure}

In Fig.~\ref{Fig:ChipIrUnfolding}, we present the plot of the ChipIr unfolded spectrum, built with the same method previously used for ROTAX, with the only difference that in this case we show the guess spectrum with the absolute normalization obtained from the MCNP6 simulation in the fast range. 
From the unfolding result, it is evident the experimental activation data point out a different fast flux spectral shape with respect to the one predicted by MCNP6 simulations.
The reason of this discrepancy is not yet understood, but from our point of view it is more likely that there is some issue in the MC simulation of high energy neutrons production and propagation through the beamline. Indeed, such simulations relies on nuclear models and cross sections affected by relatively high uncertainties.
On the other hand, since the multi-foil activation technique is based on activity measurements of which we have full control and on neutron cross sections, we can state that the unfolding results in the fast region are reliable within the same range of systematic uncertainty that can affect the cross sections of threshold reactions.
Therefore, this experimental result will be useful as a benchmark for improving the MC simulations of neutrons in the ChipIr beamline and, more generally, at the ISIS spallation source.
Beyond that, the activation data collected for this study, being obtained with the same neutron flux, could be used as a relative benchmark for the corresponding cross sections, hinting possible discrepancies in their evaluations.

\begin{table}[]
\caption{Results of the neutron flux unfolding in the ChipIr beamline}
\begin{tabular}{c|ll|ll}
\hline
Group & \multicolumn{2}{|c|}{Energy range (MeV)} & \multicolumn{2}{|c|}{Neutron Flux (cm$^{-2}$s$^{-1}$)} \\
\hline
1 &   $10^{-9}$        &   $4\times10^{-8}$   & $(1.12\pm0.16)\times10^{5}$ &   (14\%)  \\ 
2 &   $4\times10^{-8}$   &   $5\times10^{-7}$   & $(2.90\pm0.38)\times10^{5}$ &   (13\%) \\ 
3 &   $5\times10^{-7}$   &   $3\times10^{-6}$           & $(1.10\pm0.16)\times10^{5}$ &   (14\%) \\ 
4 &   $3\times10^{-6}$           &   $10^{-5}$           & $(7.64\pm0.57)\times10^{4}$ &   (7.4\%) \\ 
5 &   $10^{-5}$           &   $10^{-4}$           & $(2.29\pm0.23)\times10^{5}$ &   (10\%)  \\ 
6 &   $10^{-4}$           &   $10^{-3}$   & $(2.44\pm0.21)\times10^{5}$ &   (8.8\%) \\ 
7 &   $10^{-3}$   &   0.5      & $(3.78\pm1.52)\times10^{5}$ &   (40\%) \\ 
8 &   0.5            & 10 & $(6.14\pm1.71)\times10^{5}$ &   (28\%) \\ 
9 &   10      &   20      & $(3.39\pm0.70)\times10^{5}$ &   (21\%) \\ 
10 &   20      &   40      & $(5.70\pm1.34)\times10^{5}$ &   (24\%) \\ 
11 &   40      &   60     & $(6.08\pm1.92)\times10^{5}$ &   (32\%) \\ 
12 &   60     &  100     & $(8.40\pm1.64)\times10^{5}$ &   (20\%) \\ 
13 &  100     &   700      & $(1.70\pm0.33)\times10^{6}$ &   (20\%) \\ 
\hline        
\multicolumn{2}{c}{}&  Total     & $(6.11\pm0.44)\times10^{6}$ &   (7.2\%)  \\
\hline
\multicolumn{5}{p{0.45\textwidth}}{The uncertainties are statistical ones only, obtained by propagating the activation rates and the cross section uncertainties. In the last column we report the corresponding relative uncertainties.}\\
\end{tabular}
\label{tab:ChipIrUnfolding}
\end{table}

\subsection{Comparison of ROTAX and ChipIr neutron flux}

In order to summarize and compare the results obtained for ROTAX and ChipIr beamlines, we present Table~\ref{tab:MacroRange} in which we sum the most anti-correlated $\phi_i$ to get the neutron flux in four macro ranges: thermal neutrons ($E<0.5\,\text{eV}$), resonance and intermediate neutrons ($0.5\,\text{eV}<E<0.5\,\text{MeV}$), fast neutrons ($0.5\,\text{MeV}<E<20\,\text{MeV}$), and very fast neutrons ($E>20\,\text{MeV}$).
For both beamlines, the uncertainty is about 6\% in the thermal range, whereas in the intermediate one it increases up to 13\% mainly because of the high uncertainty affecting the 7$^\text{th}$ group, which is poorly constrained by the activation data. 
This feature is common to the flux unfolding of both beamlines, and it is due to the lack of (n,$\gamma$) reactions with big resonances above 10~keV. However, it should be noted that in the ROTAX unfolding, thanks to the activation of more foils (namely the NaCl and ZnSe ones) and thanks to the use of Cd covers, we have been able to better constrain the flux in the 1 to 10 keV range, as compared with the ChipIr case.
We also underline that the uncertainties quoted for the fast fluxes are statistical only and do not include possible systematic uncertainties related to cross sections and simulated guess spectra.

\begin{figure}[]
\centerline{\includegraphics[width=0.49\textwidth]{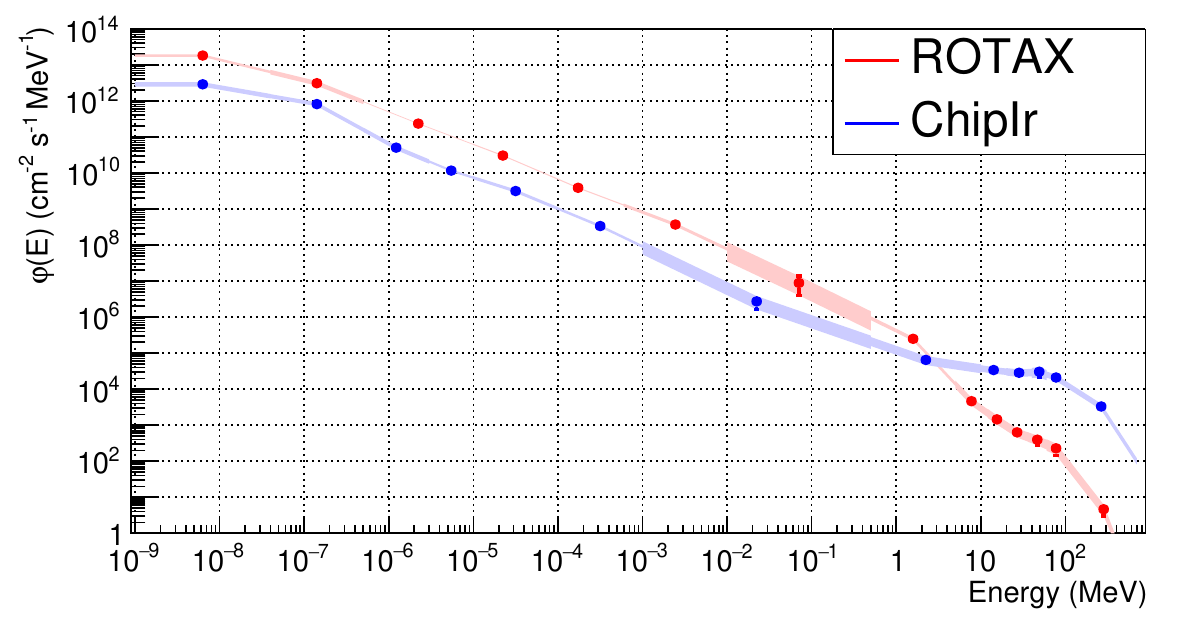}}
\caption{Continuous representation of the spectra in the ROTAX and ChipIr beamlines, obtained through a logarithmic interpolation of the differential flux values at the center of each energy group.}
\label{Fig:PlotCombined}
\end{figure}

Finally, in Fig.~\ref{Fig:PlotCombined}, we show a logarithmic interpolation of the differential flux values at the center of each bin to get a continuous representation of the spectra in both beamlines. 
Even if this plot does not exactly preserve the neutron flux intensities (due to the interpolation process), it is however useful to compare the neutron fluxes measured in the ROTAX and ChipIr beamlines. 
Particularly, it highlights that in the thermal and intermediate range, ROTAX has a flux intensity about 10 times higher with respect to ChipIr, whereas in the range of very fast neutrons, the ROTAX versus ChipIr flux ratio is $\sim 10^{-2}$.

\begin{table}[]
\caption{ROTAX and ChipIr neutron flux in four macro ranges}
\begin{tabular}{l|ll|ll}
\hline
 \multicolumn{1}{c|}{Energy range (MeV)} & \multicolumn{4}{|c}{Neutron Flux ($10^6\,\text{cm}^{-2}\text{s}^{-1}$)} \\
 \multicolumn{1}{c}{Min -- Max}  & \multicolumn{2}{|c|}{ROTAX} & \multicolumn{2}{|c}{ChipIr} \\
\hline
   $10^{-9}$ -- $5\times10^{-7}$ & $1.8\pm0.1$ & (6\%) & $0.40\pm0.03$ & (6\%) \\ 
   $5\times10^{-7}$ -- 0.5       & $9.3\pm1.1$ & (12\%) & $1.04\pm0.14$ & (13\%) \\ 
   0.5 -- 20                     & $1.16\pm0.14$ & (12\%) & $0.95\pm0.16$ & (17\%) \\ 
   20 -- 800 (700)               & $0.032\pm0.004$ & (12\%) & $3.7\pm0.4$ & (11\%) \\ 
\hline
\multicolumn{5}{p{0.45\textwidth}}{The fluxes in the four macro ranges are obtained by summing the corresponding $\phi_i$ results reported in \textsc{Tables} \ref{tab:RotaxUnfolding} and \ref{tab:ChipIrUnfolding}, by keeping into account their correlations. The uncertainties are statistical ones only, obtained by propagating the activation rates and the cross section uncertainties. The upper bound for ChipIr is 700~MeV because activation measurements have been performed with the synchrotron accelerating protons at 700~MeV (instead of 800~MeV as usual).}\\
\end{tabular}
\label{tab:MacroRange}
\end{table}

\section{Conclusions}

In this work, we showed how multi-foil activation analysis was successfully applied to get relatively accurate and precise characterization of the neutron field in two spallation beamlines used for radiation damage tests.
The method and the results are of general interest for the characterization of beamlines at advanced neutron sources.
Particularly, we presented a complete set of reactions that can be used to extract independent information from thermal to high energy neutrons in both moderated and unmoderated beamlines.
The results of these measurements will be used as a benchmark for validating the MC simulations of the neutrons produced at the ISIS spallation source, and will serve as a reference for future neutron flux measurements performed with other experimental techniques and for dose calculations in radiation damage tests.

\section*{Acknowledgment}

This work makes use of the \textit{Arby} software for \textsc{Geant4} based Monte Carlo simulations, that has been developed in the framework of the Milano - Bicocca R\&D activities and that is maintained by O. Cremonesi and S. Pozzi.

\bibliographystyle{IEEEtran}
\bibliography{Bibliography}

\end{document}